\documentclass[11pt]{article}
\usepackage{verbatim,amsmath,eufrak}
\usepackage{cite} 

\renewcommand{\arraystretch}{1.2}
\makeatletter
\newdimen\normalarrayskip              
\newdimen\minarrayskip                 
\normalarrayskip\baselineskip
\minarrayskip\jot
\newif\ifold             \oldtrue            \def\new{\oldfalse}
\def\arraymode{\ifold\relax\else\displaystyle\fi} 
\def\eqnumphantom{\phantom{(\theequation)}}     
\def\@arrayskip{\ifold\baselineskip\z@\lineskip\z@
     \else
     \baselineskip\minarrayskip\lineskip2\minarrayskip\fi}
\def\@arrayclassz{\ifcase \@lastchclass \@acolampacol \or
\@ampacol \or \or \or \@addamp \or
   \@acolampacol \or \@firstampfalse \@acol \fi
\edef\@preamble{\@preamble
  \ifcase \@chnum
     \hfil$\relax\arraymode\@sharp$\hfil
     \or $\relax\arraymode\@sharp$\hfil
     \or \hfil$\relax\arraymode\@sharp$\fi}}
\def\@array[#1]#2{\setbox\@arstrutbox=\hbox{\vrule
     height\arraystretch \ht\strutbox
     depth\arraystretch \dp\strutbox
     width\z@}\@mkpream{#2}\edef\@preamble{\halign
\noexpand\@halignto
\bgroup \tabskip\z@ \@arstrut \@preamble \tabskip\z@ \cr}%
\let\@startpbox\@@startpbox \let\@endpbox\@@endpbox
  \if #1t\vtop \else \if#1b\vbox \else \vcenter \fi\fi
  \bgroup \let\par\relax
  \let\@sharp##\let\protect\relax
  \@arrayskip\@preamble}
\def\eqnarray{\stepcounter{equation}%
              \let\@currentlabel=\theequation
              \global\@eqnswtrue
              \global\@eqcnt\z@
              \tabskip\@centering
              \let\\=\@eqncr
 \halign to \displaywidth\bgroup
    \eqnumphantom\@eqnsel\hskip\@centering
    $\displaystyle \tabskip\z@ {##}$%
    \global\@eqcnt\@ne \hskip 2\arraycolsep
         $\displaystyle\arraymode{##}$\hfil
    \global\@eqcnt\tw@ \hskip 2\arraycolsep
         $\displaystyle\tabskip\z@{##}$\hfil
         \tabskip\@centering
    &{##}\tabskip\z@\cr}
\begingroup\ifx\undefined\newsymbol \else\def\input#1 {\endgroup}\fi

\newcounter{app}

\def\app{\setcounter{equation}{0}
\def\theequation{A\Roman{app}.\arabic{equation}}\par
   \addvspace{4ex}
   \@afterindentfalse
  \secdef\@app\@dapp}
\newcommand\@app{\@startsection {app}{1}{0ex}%
                                   {-3.5ex \@plus -1ex \@minus -.2ex}%
                                   {2.3ex \@plus.2ex}%
                                   {\normalfont\Large\bf}}

\def\@dapp#1{%
{\parindent \z@ \raggedright  \bf #1}\par\nobreak}
\def\l@app#1#2{\ifnum \c@tocdepth >\z@
    \addpenalty\@secpenalty
    \addvspace{1.0em \@plus\p@}%
    \setlength\@tempdima{8.5em}%
    \begingroup
      \parindent \z@ \rightskip \@pnumwidth
      \parfillskip -\@pnumwidth
      \leavevmode \bfseries
      \advance\leftskip\@tempdima
      \hskip -\leftskip
      #1\nobreak\hfil \nobreak\hb@xt@\@pnumwidth{\hss #2}\par
    \endgroup\fi}
\newcounter{sapp}[app]

\def\sapp{\def\theequation{A\arabic{app}.\arabic{equation}}\par
   \@afterindentfalse
  \secdef\@sapp\@dsapp}
\newcommand\@sapp{\@startsection{sapp}{2}{\z@}%
                                     {-3.25ex\@plus -1ex \@minus -.2ex}%
                                     {1.5ex \@plus .2ex}%
                                     {\normalfont\large\bfseries}}

\def\@dsapp#1{%
{\parindent \z@ \raggedright  \bf #1}\par\nobreak}
\newcommand{\l@sapp}{\@dottedtocline{2}{1.5em}{3em}}
\def\draft{\oddsidemargin -.5truein
        \def\@oddfoot{\sl preliminary draft \hfil
        \rm\thepage\hfil\sl\today\quad\militarytime}
        \let\@evenfoot\@oddfoot \overfullrule 3pt
        \let\label=\draftlabel
        \let\marginnote=\draftmarginnote
   \def\@eqnnum{(\theequation)\rlap{\kern\marginparsep\tt\@eqnlabel}%
\global\let\@eqnlabel\@vacuum}  }

\def\be{\begin{eqnarray}}
\def\ee{\end{eqnarray}}

\def\p{\partial}

\def\beq{\begin{equation}}
\def\eeq{\end{equation}}
\def\ba{\beq\new\begin{array}{c}}
\def\ea{\end{array}\eeq}
\def\be{\ba}
\def\ee{\ea}

\def\Tr{{\rm Tr}\,}

\def\diag{{\rm diag}\,}

\newfont{\Bbbb}{msbm7 scaled 1\@ptsize00}

\newcommand{\z}{\raise-1pt\hbox{$\mbox{\Bbbb Z}$}}
\def\normordboson{ {\scriptstyle {{*}\atop{*}}} }

\newfont{\alef}{msbm10 at 11pt}
\newfont {\goth}{eufm10 at 11pt}
\def\mathbb#1{\hbox{{\alef #1}}}

\let\@@savethanks\thanks
\def\thanks#1{\gdef\thefootnote{\alph{footnote}}\@@savethanks{#1}}

\makeatletter
\g@addto@macro \normalsize {%
 \setlength\abovedisplayskip{14pt plus 3pt minus 3pt}%
 \setlength\belowdisplayskip{14pt plus 3pt minus 3pt}%
 \setlength\abovedisplayshortskip{11pt plus 3pt minus 3pt}%
 \setlength\belowdisplayshortskip{11pt plus 3pt minus 3pt}%
}
\makeatother

\bigskip
\bigskip

\title{
\bigskip
{\bf
Open intersection numbers, Kontsevich--Penner model and cut-and-join operators} \vspace{.5cm}}
\author{{\bf Alexander Alexandrov}\thanks{E-mail:  {\tt alexandrovsash at gmail.com}}
\date{ } \\
{\small {\it  Mathematics Institute, University of Freiburg,
Eckerstrasse 1, 79104 Freiburg, Germany \&}}\\
{\small {\it 
ITEP,
Bolshaya Cheremushkinskaya 25, 117218 Moscow, Russia}}\\
}

\begin{document}

\setcounter{footnote}{0}

\setcounter{tocdepth}{3}

\maketitle

\vspace{-8.0cm}

\begin{center}
\hfill ITEP/TH-40/14
\end{center}

\vspace{6.5cm}
\begin{abstract} 
We continue our investigation of the Kontsevich--Penner model, which describes intersection theory on moduli spaces both for open and closed curves. In particular, we show how Buryak's residue formula, which connects two generating functions of intersection numbers, appears in the general context of matrix models and tau-functions. This allows us to prove that the Kontsevich--Penner matrix integral indeed describes open intersection numbers. For arbitrary $N$ we show that the string and dilaton equations completely specify the solution of the KP hierarchy. We derive a complete family of the Virasoro and W-constraints, and using these constraints, we construct the cut-and-join operators. The case $N=1$, corresponding to open intersection numbers, is particularly interesting: for this case we obtain two different families of the Virasoro constraints, so that the difference between them describes the dependence of the tau-function on even times.
\end{abstract}
\bigskip

{Keywords: enumerative geometry, matrix models, tau-functions, KP hierarchy, Virasoro constraints, cut-and-join operator}\\

\newpage 


\def\thefootnote{\arabic{footnote}}

\section*{Introduction}
\addcontentsline{toc}{section}{Introduction}
\setcounter{equation}{0}

In our previous paper \cite{Aopen} we have shown that the Kontsevich--Penner model \cite{MMS,BH,Chekhov,GKM,Penner} is directly related to the intersections on the moduli spaces. Namely, we claimed that in addition to the well-known description of the intersections on the moduli spaces of the closed Riemann surfaces \cite{Konts,Witten,oneq} it also describes intersection theory on the moduli spaces of the Riemann surfaces with boundary. This intersection theory has been recently constructed in genus zero by R. Pandharipande, J. Solomon, R. Tessler and  further investigated by A. Buryak \cite{PST,Buryak,Buryak2} (see also \cite{Ke, Bertola}). In particular, a conjectural description of all descendants on the boundary in higher genera was introduced. 

The Kontsevich--Penner matrix model 
\be\label{Mint}
\tau_N=\det(\Lambda)^N {\mathcal C}^{-1} \int \left[d \Phi\right]\exp\left(-{\Tr\left(\frac{\Phi^3}{3!}-\frac{\Lambda^2 \Phi}{2}+N\log \Phi\right)}\right)
\ee
over $M\times M$ Hermitian matrices for $N=0$ coincides with the famous Kontsevich matrix model, which is known to describe the intersections on the moduli spaces of closed Riemann surfaces. In this paper we prove that for $N=1$ this integral indeed can be identified with the generating function of open intersection numbers conjectured by R. Pandharipande, J. Solomon, R. Tessler, and A. Buryak. For this purpose, in particular, we prove that Buryak's residue formula \cite{Buryak2}, which describes a relation between open and closed intersection numbers, follows from the matrix integral representation (\ref{Mint}). Moreover, we show how a generalization of Buryak's formula appears in the general context of the Grassmannian description of the KP/Toda-type integrable hierarchies \cite{Sato,Segal}. 

In this paper we also draw attention to the properties of the tau-function (\ref{Mint}) for general $N$. 
Using the Sato Grassmannian description we derive the full family of the Virasoro and W-constraints, which completely specify the partition function of the Kontsevich-Penner model for arbitrary $N$. In particular, (\ref{Mint}) satisfies the string equation
\be\label{stringint}
\left(\sum_{k=3}^\infty k\, t_k\, \frac{\p}{\p t_{k-2}}+\frac{t_1^2}{2}-\frac{\p}{\p t_1}+ 2\,N\, t_2\right)\tau_N=0,
\ee
and the dilaton equation
\be\label{dilatonint}
\left(\sum_{k=1}^\infty k\, t_k\, \frac{\p}{\p t_k}-\frac{\p}{\p t_3}+\frac{1}{8}+\frac{3N^2}{2}\right)\tau_N=0.
\ee
 Contrary to the constraints for the generalized Kontsevich model with the monomial potential \cite{Fukuma,MMS,Adler} our constraints for $N\neq 0$ in general do not belong to the $W_{1+\infty}$ algebra of symmetries of the integrable hierarchy. Obtained constraints allowed us to construct the cut-and-join type operator, which yields an explicit expression for the tau-function (\ref{Mint}).\footnote{Actually, we constructed a family of the cut-and-join type operators. It is not yet clear to us which representative of this family corresponds to the geometric cut-and-join analysis (if any). }

The coefficients of the series expansion of (\ref{Mint}) depend of the parameter $N$ in a relatively simple way. Namely, they are polynomials in $N$. This property allows us to consider $N$ as a continuous parameter. As we have already seen, at least for two values of $N$ the Kontsevich--Penner matrix integral gives the solutions to interesting problems of enumerative geometry. However, the properties of the generating functions for these two cases are quite different. The case $N=0$, which describes the Kontsevich--Witten tau-function of the KdV hierarchy, is very well studied. In particular, to completely specify the generating function in this case we do not need higher W-constraints, and the cut-and-join operator can be derived from the the Virasoro constraints \cite{KontsCaJ}. 
It appears that in the case of open intersection numbers ($N=1$) we have a one-parametric family of the Virasoro constraints. An operator associated with the parameter describes the dependence of the tau-function on even times
\be\label{evenint}
(k-1)\frac{\p}{\p t_{2k}}\tau_1=\sum_{j=1}^{k-1}\frac{\p^2}{\p t_{2j}\p t_{2(k-j)}}\tau_1.
\ee
This relation for the generating function of the open intersection numbers was established in \cite{Buryak2}. We claim that for a positive integer $N$ the tau-function (\ref{Mint}) is also related to interesting enumerative geometry and topological string theory models. In this paper we describe in some details the case $N=2$. For this case we have two families of the cubic W-operators. The difference between them describes a dependence of the even times $t_{2k}$ for $k>2$ and yields an analog of the relation (\ref{evenint}).  

It is well known that there exists a unique KdV tau-function, satisfying the string equation, namely, the Kontsevich--Witten tau-function \cite{Fukuma,DVV}. We found an analogous description for the Kontsevich--Penner model (\ref{Mint}). Namely, we prove that for arbitrary $N$ there is a unique tau-function of the KP hierarchy, satisfying both the string equation (\ref{stringint}) and dilaton equation (\ref{dilatonint}). 


The present paper is organized as follows. In Section \ref{S1} we briefly remind the reader the action of the $w_{1+\infty}$ algebra of symmetries on Sato's Grassmannian and show, how one can describe the acton of some simple operators from the universal enveloping algebra of $W_{1+\infty}$ on the tau-functions. In Section \ref{S2} we prove that the tau-function (\ref{Mint}) for $N=1$ is given by Buryak's formula, thus proving the matrix integral representation of the conjectural generating function of open intersection numbers. Section \ref{S3} contains the derivation of the finite number of Virasoro and cubic W-constraints for general $N$, which follow from the existence of the Kac--Schwars operators and belong to the $W_{1+\infty}$ algebra. In Section \ref{S4} we derive the complete (infinite) family of the Virasoro and cubic W-constraints (that, in general do not belong to $W_{1+\infty}$), which allow us to construct the cut-and-join operator in Section \ref{S5}. Sections \ref{S6} and \ref{S7} are devoted to the case $N=1$, which corresponds to the open intersection numbers. In Section \ref{S8} we briefly describe the tau-function (\ref{Mint}) for integer $N>1$, in particular, we investigate the dependence on the even times for the next potentially interesting case ($N=2$). In Appendix A we give the first terms of the series expansion of the tau-function $\tau_N$ and the corresponding free energy.

\section{$W_{1+\infty}$ algebra and the Sato Grassmannian}\label{S1}

In this section we give a brief reminder of some important properties of the algebra $w_{1+\infty}$ and its central extension, the algebra $W_{1+\infty}$. They describe the symmetries of the KP integrable hierarchy and play a central role in our construction. For more details see, i.e., \cite{Sato,Segal,Fukuma, AHeis} and references therein.

The KP hierarchy can be described by the bilinear identity, satisfied by the tau-function $\tau({\bf t})$, namely  
\begin{equation}\label{bi1}
\oint_{{\infty}} e^{\xi ({\bf t}-{\bf t'},z)}
\,\tau ({\bf t}-[z^{-1}])\,\tau ({\bf t'}+[z^{-1}])dz =0,
\end{equation}
where $\xi({\bf t},z)=\sum_{k=1}^\infty t_k z^{k}$ and we use the standard notation
\be\label{shiftedt}
{\bf t} \pm \left[z^{-1}\right]=\left\{t_1\pm\frac{1}{z},t_2\pm\frac{1}{2z^2},t_3\pm\frac{1}{3z^3},\dots\right\}.
\ee

From the free fermion description of the KP hierarchy it immediately follows that the operators
\be
\widehat{W}^{(m+1)}(z)=\normordboson 
\left(\widehat{J}(z)+\p_z \right)^m \widehat{J}(z)\normordboson
\ee
correspond to the bilinear combinations of fermions and span the algebra $W_{1+\infty}$ of symmetries of the KP hierarchy.\footnote{Omitting some details, one can say that a group element $e^{\widehat{W}}$, where $\widehat{W}\in W_{1+\infty}$, maps a tau function $\tau$ to another tau-function $e^{\widehat{W}}\tau$. 
} 
Here $\widehat{J}(z)$ is the so-called bosonic current
\be
\widehat{J}(z)= \sum_{m \in \z} \frac{\widehat{J}_m}{z^{m+1}},
\ee
where
\be
\widehat{J}_k =
\begin{cases}
\displaystyle{\frac{\p}{\p t_k} \,\,\,\,\,\,\,\,\,\,\,\, \mathrm{for} \quad k>0},\\[10pt]
\displaystyle{0}\,\,\,\,\,\,\,\,\,\,\,\,\,\,\,\,\,\,\, \mathrm{for} \quad k=0,\\[10pt]
\displaystyle{-kt_{-k} \,\,\,\,\,\mathrm{for} \quad k<0.}
\end{cases}
\ee
The normal ordering for bosonic operators $\normordboson\dots\normordboson$ puts all operators $\widehat{J}_k$ with positive $k$ to the right of all $\widehat{J}_k$ with negative $k$.

The Virasoro subalgebra of $W_{1+\infty}$ is generated by the operators, bilinear in $\widehat{J}_k$
\be
\frac{1}{2}\normordboson \widehat{J}(z)^2\normordboson = \sum_{m\in \z} \frac{\widehat{L}_m}{z^{m+2}},
\ee
namely it is spanned by the operators
\begin{equation}
\label{virfull}
\widehat{L}_m=\frac{1}{2} \sum_{a+b=-m}a b t_a t_b+ \sum_{k=1}^\infty k t_k \frac{\p}{\p t_{k+m}}+\frac{1}{2} \sum_{a+b=m} \frac{\p^2}{\p t_a \p t_b}.
\end{equation}
The operators from the $W^{(3)}$ algebra,
\begin{multline}
\widehat{M}_k=\frac{1}{3} \sum_{a+b+c=k} \normordboson \widehat{J}_a \widehat{J}_b \widehat{J}_c \normordboson=
\frac{1}{3}\sum_{a+b+c=-k}a\, b\, c\, t_a\, t_b\, t_c+\sum_{c-a-b=k}a\, b\, t_a\, t_b\, \frac{\p}{\p t_{c}}\\
+\sum_{b+c-a=k}a\, t_{a}\frac{\p^2}{\p t_b\p t_c}+\frac{1}{3}\sum_{a+b+c=k}\frac{\p^3}{\p t_a \p t_b \p t_c},
\end{multline}
are generated by
\be
\frac{1}{3}\normordboson \widehat{J}(z)^3\normordboson=\sum_{m\in \z}\frac{\widehat{M}_m}{z^{m+3}}.
\ee
The operators $\widehat{J}_k$, $\widehat{L}_k$, and $\widehat{M}_k$ satisfy the following commutation relations
\be
\left[\widehat{J}_k,\widehat{J}_m\right]=k\, \delta_{k,-m},\\
\left[\widehat{J}_k,\widehat{L}_m\right]=k\widehat{J}_{k+m},\\
\left[\widehat{L}_k,\widehat{L}_m\right]=(k-m)\widehat{L}_{k+m}+\frac{1}{12}k(k^2-1)\delta_{k,-m},\\
\left[\widehat{L}_k,\widehat{M}_m\right]=(2k-m)\widehat{M}_{k+m}+\frac{1}{6}k(k^2-1)\widehat{J}_{k+m},\\
\left[\widehat{J}_k,\widehat{M}_m\right]=2k\,\widehat{L}_{k+m}.\\
\ee
A commutator of two operators from $W^{(3)}$ contains the terms of fourth power of the current components $\widehat{J}_m$, so it can not be represented as a linear combination of $\widehat{J}_k$, $\widehat{L}_k$, and $\widehat{M}_k$.

The description of the integrable hierarchies in terms of the Grassmannian \cite{Sato,Segal} allows us to work with the operators from the algebra $w_{1+\infty}$ (the differential operators in one variable, which describe diffeomorphisms of the circle) instead of the operators from $W_{1+\infty}$. This significantly simplifies the calculations in some cases. In this paper we consider the tau-functions, given by matrix models of the Kontsevich type. Thus, we use the following Miwa parametrization
\be\label{Miwapar}
t_k=\frac{1}{k}\Tr Z^{-k}
\ee
for a diagonal matrix $Z=\diag(z_1,z_2,\dots,z_M)$.
A tau-function of the KP hierarchy in this parametrization is given by
\be\label{Miwatau}
\tau\left(\left[Z^{-1}\right]\right)=\frac{{\det_{i,j=1}^M \Phi_i (z_j)}}{\Delta(z)},
\ee
where $\Delta(z)$ is the Vandermonde determinant and we use a natural generalization of the notation (\ref{shiftedt}), $\left[Z^{-1}\right]=\left[z_1^{-1}\right]+\left[z_2^{-1}\right]+\dots+\left[z_M^{-1}\right]$.\footnote{In \cite{Aopen} the same Miwa parametrization was denoted by $\left[Z\right]$.}  Here 
\be
\Phi_i (z)=z^{i-1}+\sum_{k=-\infty}^{i-2}\Phi_{ik} z^k 
\ee
are known as the basis vectors and define a point of the Sato Grassmannian. Let us denote $\left\{\Phi\right\}=\{\Phi_1,\Phi_2,\Phi_3,\dots\}$. We call an operator $a \in w_{1+\infty}$ the Kac--Schwarz (KS) operator for the tau-function $\tau$ if for the corresponding point of the Sato Grassmannian we have
\be\label{KScond}
a\, \left\{\Phi\right\} \subset \left\{\Phi\right\}.
\ee 

For the parametrization (\ref{Miwapar}) a relation between the algebras $w_{1+\infty}$ and  $W_{1+\infty}$ is as follows. With any operator $a\in w_{1+\infty}$ we identify an operator $\widehat{Y}_a\in W_{1+\infty}$ such that the operators $\left(-\p_z \right)^m z^{-k}$ (which span the algebra $w_{1+\infty}$) are identified  \cite{AHeis} with:
\be\label{Mastid}
\widehat{Y}_{\left(-\p_z \right)^m z^{-k}}=\mbox{res}_z\left(z^{-k}\normordboson \frac{(\widehat{J}(z)+\p_z)^{m}}{m+1} \widehat{J}(z)\normordboson \right).
\ee

When the size $M$ of the auxiliary matrix $Z$ in (\ref{Miwatau}) tends to infinity, we have an infinite number of Miwa parameters: 
\be\label{taumi}
\tau\left(\left[Z^{-1}\right]\right)=\frac{{\det_{i,j=1}^\infty \Phi_i (z_j)}}{\Delta(z)}.
\ee
Then, for any operator $b\in w_{1+\infty}$ and the corresponding operator $\widehat{Y}_b\in W_{1+\infty}$ we have a family of the group elements:
\be
\left.e^{\epsilon \widehat{Y}_b} \,\tau\left({\bf t}\right)\right|_{t_k=\frac{1}{k}\Tr Z^{-k}}=\frac{\det_{i,j=1}^\infty e^{\epsilon b}\Phi_i (z_j)}{\Delta(z)}. 
\ee
Here $\epsilon$ is an arbitrary parameter and it is assumed that in the $j$th row the operator $b$ acts on the variable $z_j$.
Let us introduce a notation for the determinant in the numerator of (\ref{taumi}):
\be
{\det_{i,j=1}^\infty \Phi_i (z_j)}=\left|\Phi_1,\, \Phi_2,\, \Phi_3,\dots\right|,
\ee
where $\Phi_i$ is an infinite column
\be
\Phi_i=\left[ \begin{array}{c}  \Phi_i(z_1) \\ \Phi_i(z_2) \\ \Phi_i(z_3)\\ \vdots \end{array} \right]. 
\ee
Then, the family of the group elements, corresponding to the $\widehat{Y}_b\in W_{1+\infty}$ acts as follows:
\be
\left.e^{\epsilon\widehat{Y}_b} \,\tau\left({\bf t}\right)\right|_{t_k=\frac{1}{k}\Tr Z^{-k}}=\frac{1}{\Delta(z)} \left|e^{\epsilon b}\Phi_1,\, e^{\epsilon b}\Phi_2,\, e^{\epsilon b}\Phi_3,\dots\right|.
\ee
 The first two terms of expansion of this identity in $\epsilon$ give respectively
\be\label{detexp1}
\widehat{Y}_b \,\tau=\frac{1}{\Delta(z)}\sum_{l=1}^\infty \left|\Phi_1,\, \Phi_2,\,\dots,\,\Phi_{l-1},\,b\,\Phi_{l},\,\Phi_{l+1},\,\dots\right|,
\ee 
and
\be\label{detexp2}
\widehat{Y}_b^2 \,\tau=\frac{1}{\Delta(z)}\left(\sum_{l=1}^\infty \left|\Phi_1,\, \Phi_2,\,\dots,\,\Phi_{l-1},\,b^2\,\Phi_{l},\,\Phi_{l+1},\,\dots\right|\right.\\
\left.+2\sum_{k<l}^\infty \left|\Phi_1,\, \Phi_2,\,\dots,\,\Phi_{k-1},\,b\,\Phi_{k},\,\Phi_{k+1},\,\dots,\,\Phi_{l-1},\,b\,\Phi_{l},\,\Phi_{l+1},\,\dots \right|\right).
\ee

\section{Buryak's residue formula}\label{S2}

In this section we prove that the tau-function $\tau_1$ indeed coincides with the conjectural generating function of open intersection numbers of \cite{Buryak2}. In particular, we show that the residue formula for the generation function, proved in \cite{Buryak2}, follows from the determinant expression (\ref{Miwatau}) for the tau-functions of the KP hierarchy. Moreover, this type of relations appears to be universal for tau-functions.

Any KP tau-function can be expanded in the Schur polynomials
\be
\tau({\bf t})=\sum_{\lambda} c_{\lambda} s_{\lambda}({\bf t}).
\ee
Then, the sum restricted to Young diagrams with at most $n$ non-zero lines 
\be
\tau^{(n)}({\bf t})=\sum_{l(\lambda)\leq n} c_{\lambda} s_{\lambda}({\bf t})
\ee
is also a KP tau-function for all $n>0$. This type of tau-functions often appears in matrix integrals \cite{Orlov,AZ}.

Assume we have an expression for the tau-function in the Miwa parametrization  $\tau\left(\left[Z^{-1}\right]\right)$ for $Z=\diag(z_1,z_2,\dots,z_M)$. Then,
the orthogonality of the Schur functions allows us to find an expression for this tau-function, dependent on an infinite number of times:
\be
\tau^{(M)}({\bf{t}})=\frac{1}{ M!}\oint\dots\oint \Delta(z)\Delta(z^{-1})\exp\left(\sum_{k=1}^\infty \sum_{l=1}^M t_k  z_l^k\right)\tau\left(\left[Z^{-1}\right]\right)\prod_{j=1}^M \frac{d z_j}{2 \pi i z_j}.
\ee
On substitution of the determinant representation (\ref{Miwatau}) we obtain
\be\label{resid1}
\tau^{(M)}({\bf{t}})=\oint\dots\oint \Delta(z^{-1})\exp\left(\sum_{k=1}^\infty \sum_{l=1}^M t_k  z_l^k\right)\prod_{j=1}^M \Phi_j(z_j) \frac{d z_j}{2 \pi i z_j}.
\ee
Here we use the antisymmetry of the Vandermonde determinant. Let us extract the integral over $z_1$:
\be
\tau^{(M)}({\bf{t}})=\oint\exp\left(\sum_{k=1}^\infty  t_k z_1^k\right) Z({\bf{t}};z_1)\,\Phi_1(z_1)\frac{d z_1}{2 \pi i z_1}, 
\ee
where
\be\label{resid2}
Z({\bf{t}};z_1)=\oint\dots\oint \Delta(z^{-1})\exp\left(\sum_{k=1}^\infty \sum_{l=2}^M t_k z_l^k\right)\prod_{j=2}^M \Phi_j(z_j) \frac{d z_j}{2 \pi i z_j}\\
= \oint\dots\oint \widetilde{\Delta}(z^{-1}) \exp\left(\sum_{k=1}^\infty \sum_{l=2}^M t_k z_l^k\right)\prod_{j=2}^M\left(\frac{1}{z_j}-\frac{1}{z_1}\right) \Phi_j(z_j) \frac{d z_j}{2 \pi i z_j}\\
= \oint\dots\oint \widetilde{\Delta}(z^{-1}) \exp\left(\sum_{k=1}^\infty \sum_{l=2}^M \left(t_k-\frac{1}{kz_1^k}\right) z_l^k\right)\prod_{j=2}^M\frac{\Phi_j(z_j)}{z_j} \frac{d z_j}{2 \pi i z_j}.
\ee
Here 
\be
\widetilde{\Delta}(z^{-1})=\prod_{i<j;\,\,i,j=2..N}(z_i-z_j)
\ee
is a determinant of the $(M-1)\times(M-1)$ Vandermonde matrix. Comparing (\ref{resid1}) and (\ref{resid2}) we see that $Z({\bf{t}}; z_1)$ can be identified with the tau-function $\widetilde{\tau}^{(M-1)}({\bf t}-\left[z_1^{-1}\right])$ which corresponds to the point of the Grassmannian $\left\{z^{-1}\Phi_{2},z^{-1}\Phi_{3},z^{-1}\Phi_{4},\dots\right\}$:
\be
\tau^{(M)}({\bf{t}})=\oint\exp\left(\sum_{k=1}^\infty  t_k z_1^k\right) \widetilde{\tau}^{(M-1)}\left({\bf t}-\left[z_1^{-1}\right]\right)\,\Phi_1(z_1)\frac{d z_1}{2 \pi \,i\, z_1}. 
\ee
When $M$ tends to infinity we get a relation
\be\label{genres}
\tau({\bf{t}})=\oint\exp\left(\sum_{k=1}^\infty  t_k z^k\right) \widetilde{\tau}\left({\bf t}-\left[z^{-1}\right]\right)\,\Phi_1(z)\frac{d z}{2 \pi i\, \,z}. 
\ee

Thus, we proved the following statement: for any tau-function $\widetilde{\tau}$ and arbitrary series $\Phi_1(z)=1+O(z^{-1})$ the residue (\ref{genres}) gives a tau-function of the KP hierarchy, corresponding to the point of the Sato Grassmannian 
\be
\left\{{\Phi}_1, z\widetilde{\Phi}_1,z\widetilde{\Phi}_2,z\widetilde{\Phi}_3,\dots\right\}.
\ee
 Moreover, it is easy to see that the resulting tau-function satisfies the MKP hierarchy equation
\be
\oint_{{\infty}} e^{\xi ({\bf t}-{\bf t'},z)}\, z
\,\tau ({\bf t}-[z^{-1}])\,\widetilde{\tau}({\bf t'}+[z^{-1}])dz =0.
\ee
The relation is a particular case of a more general relation between tau-functions. Namely, in the same way it is easy to show that for any tau-functions $\widetilde{\tau}$ and $\tau^*$ the function $\tau$ defined by
\be\label{GENI}
\tau({\bf{t}})=\frac{1}{ M!}\oint\dots\oint \Delta(z)\Delta(z^{-1}) \widetilde{\tau}\left({\bf t}-\left[Z^{-1}\right]\right) \tau^*\left(\left[Z^{-1}\right]\right)\exp\left(\sum_{k=1}^\infty \sum_{l=1}^M t_k  z_l^k\right)\prod_{j=1}^M \frac{d z_j}{2 \pi i z_j}
\ee
is a tau-function. The corresponding point of the Sato Grassmannian is given by 
\be
\left\{{\Phi}_1^*,{\Phi}_2^*,\dots,{\Phi}_M^*, z^M\widetilde{\Phi}_1, z^M\widetilde{\Phi}_2, z^M\widetilde{\Phi}_3,\dots\right\}.
\ee

For the tau-function $\tau_N$ of the Kontsevich-Penner model considered in Section \ref{S3} relation (\ref{genres}) reduces to
\be
\tau_{N}({\bf{t}})=\oint\exp\left(\sum_{k=1}^\infty  t_k z^k\right) \tau_{N-1}\left({\bf t}-\left[z^{-1}\right]\right)\,\Phi_1^N(z)\frac{d z}{2 \pi i z}. 
\ee
In particular, for $N=1$ we have
\be\label{Resid}
\tau_1({\bf{t}})=\oint\exp\left(\sum_{k=1}^\infty  t_k z^k\right) \tau_{0}\left({\bf t}-\left[z^{-1}\right]\right)\,\Phi_1^1(z)\frac{d z}{2 \pi i z}. 
\ee
Since $\tau_{0}=\tau_{KW}$, the r.h.s.~coincides with the expression for the generating function of the open intersection numbers, derived by A. Buryak in \cite{Buryak2}. Thus, we proved that the conjectural generating function of open intersection numbers is given by the Kontsevich-Penner model for $N=1$
\be
\tau_o=\tau_{1}.
\ee

\section{Kac--Schwarz operators and corresponding constraints for general $N$}\label{S3}

As we have established in our previous work \cite{Aopen}, an operator
\be
a_{N}=\frac{1}{z}\frac{\p}{\p z} -\left(N+\frac{1}{2}\right)\frac{1}{z^2}+z 
\ee
is the KS operator for the tau-function, corresponding to the Kontsevich-Penner model
\be
\tau_N=\frac{\displaystyle{\int\left[d \Phi\right]\,\det\left(1+\frac{\Phi}{\Lambda}\right)^{-N}\exp\left(-{\Tr\left(\frac{\Phi^3}{3!}+\frac{\Lambda \Phi^2}{2}\right)}\right)}}{\displaystyle{\int\left[d \Phi\right]\exp\left(-{\Tr\frac{\Lambda \Phi^2}{2}}\right)}}\\
=\det(\Lambda)^N {\mathcal C}^{-1} \int \left[d \Phi\right]\exp\left(-{\Tr\left(\frac{\Phi^3}{3!}-\frac{\Lambda^2 \Phi}{2}+N\log \Phi\right)}\right).
\label{matint}
\ee
Namely, the basis vectors 
\be\label{bascon}
\Phi_k^N=z^N \Phi_{k-N}^0=\frac{z^{N+1/2}}{\sqrt{2\pi}}e^{-\frac{z^3}{3}}\int_{C} d\, y \, y^{k-N-1} \exp\left(-\frac{y^3}{3!}+\frac{y z^2}{2}\right),
\ee
with a properly chosen contour $C$, satisfy a relation
\be\label{atop}
a_{N}\Phi_i^N=\Phi_{i+1}^N.
\ee
These basis vectors have an expansion
\be\label{expbas}
\Phi_k^N
=z^{k-1}+\frac{12(2-p)^2-7}{24}z^{k-4}
+\left(
\frac{1}{8}p^4-\frac{5}{3}p^3+\frac{365}{48}p^2-\frac{55}{4}p+\frac{9241}{1152}\right)z^{k-7}\\
+ \left(\frac{1}{48}{p}^{6}-{\frac {7}{12}}\,{p}^{5}+{\frac {1225}{192}
}\,{p}^{4}-{\frac {2485}{72}}\,{p}^{3}+{\frac {
221137}{2304}}\,{p}^{2}-{\frac {73409}{576}}\,p
+ {\frac {5075225}{82944}}\right) {z}^{k-10}+O(z^{k-13}),
\ee
where  $p=k-N$. Using the integral representation (\ref{bascon}) it is easy to see that the operator of multiplication by $z^2$ acts as follows:
\be\label{z2op}
z^2\Phi_k^N= \Phi_{k+2}^N-2(k-N-1)\Phi_{k-1}^N.
\ee
This operator is not the KS operator for $N\neq0$, because
\be
z^2\Phi_1^N= \Phi_{3}^N+2N\Phi_{0}^N \notin \left\{\Phi^{N}\right\},\,\,\,\,\,\,\,\,\,\,\mathrm{for}\,\,  N\neq 0.
\ee
However, it is straightforward to check that the operators 
\be\label{simplevir}
\mathsf{l}_{-1}=-a_N,\\
\mathsf{l}_{0}=-z^2a_N+N-1,\\
\mathsf{l}_{1}=-z^4a_N+2(N-1)z^2,
\ee
are the KS operators for any $N$ \cite{Aopen}. For example, from (\ref{z2op}) we see that for $\mathsf{l}_{1}$ it is enough to check the condition (\ref{KScond}) only for $\Phi_1^N$.
A constant term in the operator $\mathsf{l}_{0}$ is chosen in such a way that the following commutation relations hold:
\be
\left[\mathsf{l}_i,\mathsf{l}_j\right]=2(i-j)\mathsf{l}_{i+j}\,\,\,\,\,\,k=-1,0,1.
\ee
The algebra $sl(2)$ generated by operators (\ref{simplevir}) can be extended to the full semi-infinite Virasoro algebra of the KS operators $\mathsf{l}_k=z^{2k+2}a_N+\dots$ with $k\geq -1$ only for $N=0$ (the KW tau-function, both $a_0$ and $z^2$ are the KS operators, so that any of their combinations is also the KS operator) and for $N=1$ (the open intersection numbers of \cite{PST,Buryak,Buryak2}, any operator $z^{2k}a_1$ for $k\geq 0$ is the KS operator).

The relation (\ref{Mastid}) allows us to find the operators from the $W_{1+\infty}$ algebra, which correspond to the operators (\ref{simplevir}):
 \be\label{strdil}
  \widehat{\mathsf{L}}_{-1}=\widehat{L}_{-2}-\frac{\p}{\p t_1}+ 2N t_2,\\
 \widehat{\mathsf{L}}_{0}=\widehat{L}_0-\frac{\p}{\p t_3}+\frac{1}{8}+\frac{3N^2}{2},\\
  \widehat{\mathsf{L}}_{1}=\widehat{L}_{2}-\frac{\p}{\p t_{5}}+3N\frac{\p}{\p t_{2}}.
 \ee
 The operators $\widehat{\mathsf{L}}_i$ also satisfy the commutation relations of the $sl(2)$ subalgebra of the Virasoro algebra
 \be
 \left[\widehat{\mathsf{L}}_{i},\widehat{\mathsf{L}}_{j}\right]=2(i-j)\,\widehat{\mathsf{L}}_{i+j},\,\,\,\,\,i,j=-1,0,1,
 \ee
so that the constrains
  \be\label{smallL}
 \widehat{\mathsf{L}}_k \tau_N =0,\,\,\,\,k=-1,0,1
 \ee
are satisfied. In what follows we call the equations with $k=-1$ and $k=0$ the string equation and the dilaton equations. \footnote{The string equation was derived by E.~Brezin and S.~Hikami in \cite{BH}. They also found a constraint, similar to our dilaton equation (but their equation is essentially different and we claim that it contains a misprint) and a constraint $\widehat{\mathsf{M}}_{-2}$ (see below).}
 
Let us show that the string and dilaton equations uniquely specify the solution of the KP hierarchy (in the same way as the string equation specifies the KW tau-function of the KdV hierarchy \cite{Fukuma,DVV}). We follow the approach of \cite{Kac}, namely, we prove that the corresponding KS operators $\mathsf{l}_{-1}$ and $\mathsf{l}_0$ completely specify a point of the Sato Grassmannian (let us note that these operators, however, do not generate the KS algebra for $\tau_N$). Indeed, the operator $\mathsf{l}_{-1}=a_N$ allows us to find all higher $(k>1)$ basis vectors $\Phi_k^N$ via (\ref{atop}) if the first basis vector is known. Thus, it remains to show that the first basis vector is completely defined by the KS operators $\mathsf{l}_{-1}$ and $\mathsf{l}_0$. Indeed, from the definition of the KS operators, it follows that the series $\mathsf{l}_0 \Phi_1^N=z^3+\dots$ should be a combination of the basis vectors:
\be\label{maste}
\mathsf{l}_0 \Phi_1^N=\left(\sum_{k=1}^3 \alpha_k a_N^k\right) \Phi_1^N.
\ee
for some constant $\alpha_k$. On substitution of the anzats 
\be\label{firstgen}
\Phi_1^N=1+\sum_{k=1}^\infty b_k z^{-k}
\ee
into this equation, we immediately obtain an expression for the coefficients $\alpha_k$:
\be\label{masteqq}
\left(a_N^3-z^2a_N+2(N-1)\right)\Phi_1^N=0.
\ee 
This equation has a unique solution of the form (\ref{firstgen}):
\be
\Phi_1^N=1+\left({\frac {5}{24}}+N+\frac{1}{2}\,{N}^{2}\right)z^{-3}+\left({\frac {385}{1152}}+{\frac {73}{24}}\,N+{\frac {161}{48}}\,{N}^{2}+\frac{7}{6}\,{N}^{3}+\frac{1}{8}\,{N}^{4}\right)z^{-6}\\
+\left({\frac {85085}{82944}}+{\frac {6259}{384}}N+{\frac {58057}{2304}}{N}^
{2}+{\frac {2075}{144}}{N}^{3}+{\frac {725}{192}}{N}^{4}+{\frac {11}{
24}}{N}^{5}+{\frac {1}{48}}{N}^{6}\right)z^{-9}+O(z^{-12}).
\ee
Thus, there is a unique KP tau-function, satisfying the string (\ref{stringint}) and dilaton (\ref{dilatonint}) equations. Equation (\ref{masteqq}) can be considered as a version of the quantum spectral curve for the Kontsevich--Penner model.

For arbitrary $N$ the operators
\be\label{smallmoper}
\mathsf{m}_{-2}=a_N^2,\\
\mathsf{m}_{-1}=z^2a_N^2-(N-2)a_N,\\
\mathsf{m}_{0}=z^4a_N^2-2(N-2)z^2a_N+\frac{2}{3}(N-1)(N-2),\\
\mathsf{m}_{1}=z^6a_N^2-3(N-2)z^4a_N+2(N-1)(N-2)z^2,\\
\mathsf{m}_{2}=z^8a_N^2-4(N-2)z^6a_N+4(N-1)(N-2)z^4,\\
\ee
are the KS operators. Of course, these operators are not unique KS operators with the leading terms $z^{2k-4}a_N^2$. Namely, one can add to them a combination of the operators (\ref{simplevir}) and a constant. Our choice corresponds to the commutation relations
\be
\left[\mathsf{l}_j,\mathsf{m}_k\right]=2\,(2j-k)\mathsf{m}_{j+k}.
\ee
The correspondence (\ref{Mastid}) for the operators (\ref{smallmoper}) yields
\be
\widehat{\mathsf{M}}_{-2}=\widehat{M}_{-4}-2\widehat{L}_{-1}+2N\left(\widehat{L}_{-4}-t_1\right)+\frac{\p}{\p t_2}+\left(4N^2+1\right)t_4,\\
\widehat{\mathsf{M}}_{-1}=\widehat{M}_{-2}-2\widehat{L}_{1}+3N\left(\widehat{L}_{-2}-\frac{\p}{\p t_1}\right)+\frac{\p}{\p t_4}+\left(4N^2+\frac{1}{2}\right)t_2,\\
\widehat{\mathsf{M}}_{0}=\widehat{M}_{0}-2\widehat{L}_{3}+4N\left(\widehat{L}_{0}-\frac{\p}{\p t_3}\right)+\frac{\p}{\p t_6}+2\left(N^2+\frac{1}{4}\right)N,\\
\widehat{\mathsf{M}}_{1}=\widehat{M}_{2}-2\widehat{L}_{5}+5N\left(\widehat{L}_{2}-\frac{\p}{\p t_5}\right)+\frac{\p}{\p t_8}+\left(6N^2+\frac{1}{4}\right)\frac{\p}{\p t_2},\\
\widehat{\mathsf{M}}_{2}=\widehat{M}_{4}-2\widehat{L}_{7}+6N\left(\widehat{L}_{4}-\frac{\p}{\p t_7}\right)+\frac{\p}{\p t_{10}}+\left(9N^2+\frac{1}{4}\right)\frac{\p}{\p t_4},\
\ee
so that in general we can write
\be
\widehat{\mathsf{M}}_{k}=\widehat{M}_{2k}-2\widehat{L}_{2k+3}+\widehat{J}_{2k+6}+\left(3(k+1)N^2+\frac{1}{4}\right)\widehat{J}_{2k}\\
+(k+4)N\left(\widehat{L}_{2k}-\widehat{J}_{2k+3}\right)+2\left(N^2+\frac{1}{4}\right)N\delta_{k,0}
+4\,N^2t_2\delta_{k,-1}+16\,N^2t_4\delta_{k,-2}.
\ee

For $k=-1,0,1$ and $m=-2,-1,0,1,2$ we have the following commutation relations
\be\label{smallcom}
\left[\widehat{\mathsf{L}}_{k},\widehat{\mathsf{M}}_{l}\right]=2\,(2k-l)\widehat{\mathsf{M}}_{k+l},
\ee
thus
  \be\label{smallM}
 \widehat{\mathsf{M}}_k \,\tau_N =0,\,\,\,\,k=-2,-1,0,1,2.
 \ee
However, equations (\ref{smallL}) and (\ref{smallM}) without referring to the integrability have more then one solution. 
In the next section we will construct an infinite family of constraints, which completely specify partition function of the Kontsevich--Penner model. These operators, in general, do not correspond to any KS operators, thus, they do not belong to the algebra $W_{1+\infty}$. However, it is possible to construct an infinite number of the independent constraints given by the operators from $W_{1+\infty}$, which would correspond to the operators from $w_{1+\infty}$ with higher powers of $a_N$ and would completely specify the generating function.

\section{Higher constraints for general $N$}\label{S4} 

The Virasoro and W-constraints for the generalized Kontsevich model can be obtained by standard matrix model techniques, namely, 
by variation of the matrix integral.\footnote{ Part of the corresponding calculation, which, however, is not enough to find the full algebra of constraints, is given in \cite{BH}.} For the Kontsevich--Penner model (\ref{matint}) with arbitrary $N\neq0$ the calculations are rather cumbersome. The reason is that for general $N$ the partition function satisfies the third order equation \cite{BH,MMS}
\be
\left(\left(\frac{1}{\Lambda}\frac{\p}{\p \Lambda^{tr}}\right)^3-\Lambda^2\left(\frac{1}{\Lambda}\frac{\p}{\p \Lambda^{tr}}\right)+2(N-M) \right)  \frac{\det(\Lambda)^N}{\mathcal C}\,\tau_N\left(\left[\Lambda^{-1}\right]\right)=0,
\ee
while for $N=0$ it can be reduced to the second order equation
\be
\left(\left(\frac{1}{\Lambda}\frac{\p}{\p \Lambda^{tr}}\right)^2-\Lambda^2\right) {\mathcal C}^{-1}\, \tau_{KW}\left(\left[\Lambda^{-1}\right]\right)=0.
\ee
In this sense the derivation of the constraints for the Kontsevich--Penner model (\ref{matint}) is of the same level of complexity as the calculations for the generalized Kontsevich model with the quartic potential, performed in \cite{Mikhi}.
 
This is why we take a different route and develop here a new approach based on the correspondence  (\ref{Mastid}) between the operators from $W_{1+\infty}$ and $w_{1+\infty}$ . Let us show that the operator
 \be
 \widehat{\mathsf{L}}_{2}=\widehat{L}_{4}-\frac{\p}{\p t_{7}}+3N\frac{\p}{\p t_{4}}+\frac{\p^2}{\p t_2^2}
 \ee
annihilates the tau-function (\ref{matint}).  This operator, because of the term $\frac{\p^2}{\p t_2^2}$, does not belong to the $W_{1+\infty}$ algebra, thus, it does not directly correspond to any KS operator.

First of all, let us consider an operator 
\be
p=-z^6 a_N+(2N-3)z^4,
\ee
which, via the identification (\ref{Mastid}), corresponds to the operator
\be
\widehat{Y}_p=\widehat{L}_{4}-\frac{\p}{\p t_{7}}+3N\frac{\p}{\p t_{4}}.
 \ee
 This operator belongs to the $W_{1+\infty}$ algebra, thus, its action can be easily considered on the level of the basis vectors. Indeed, from (\ref{atop}) and (\ref{z2op}) it immediately follows that only two terms in the r.h.s. of (\ref{detexp1}) survives:
 \be\label{High1}
 \widehat{Y}_p \,\tau_N=\frac{1}{\Delta(z)}\left( \left|p\,\Phi_1^N,\, \Phi_2^N,\,\Phi_3^N,\,\dots\right|+ \left|\Phi_1^N,\, p\,\Phi_2^N,\,\Phi_3^N,\,\dots\right|\right)\\
 =\frac{1}{\Delta(z)}\left( -4N(N+1)\left|\Phi_{-1}^N,\, \Phi_2^N,\,\Phi_3^N,\,\dots\right|+ 4N(N-1)\left|\Phi_1^N,\, \,\Phi_0^N,\,\Phi_3^N,\,\dots\right|\right).
\ee 
For the operator $q=z^2$ the corresponding operator from the algebra $W_{1+\infty}$ is $\widehat{Y}_q=\frac{\p}{\p t_2}$ and from (\ref{detexp2}) it follows that
\be\label{High2}
\frac{\p^2}{\p t_2^2} \,\tau_N=\frac{1}{\Delta(z)}\left(\sum_{l=1}^\infty \left|\Phi_1^N,\, \Phi_2^N,\,\dots,\,\Phi_{l-1}^N,\,z^4\,\Phi_{l}^N,\,\Phi_{l+1}^N,\,\dots\right|\right.\\
\left.+2\sum_{k<l}^\infty \left|\Phi_1^N,\, \Phi_2^N,\,\dots,\,\Phi_{k-1}^N,\,z^2\,\Phi_{k}^N,\,\Phi_{k+1}^N,\,\dots,\,\Phi_{l-1}^N,\,z^2\,\Phi_{l}^N,\,\Phi_{l+1}^N,\,\dots \right|\right).
\ee 
 Again, in the first sum only terms with $l=1$ and $l=2$ survive
 \be\label{High3}
\sum_{l=1}^\infty \left|\Phi_1^N,\, \Phi_2^N,\,\dots,\,\Phi_{l-1}^N,\,z^4\,\Phi_{l}^N,\,\Phi_{l+1}^N,\,\dots\right|\\
= 4N(N+1)\left|\Phi_{-1}^N,\, \Phi_2^N,\,\Phi_3^N,\,\dots \right|+4N(N-1)\left|\Phi_1^N,\, \,\Phi_0^N,\,\Phi_3^N,\,\dots\right|,
 \ee
 while in the double sum only a term with $k=1$ and $l=2$  survives
 \be\label{High4}
 2\sum_{k<l}^\infty \left|\Phi_1^N,\, \Phi_2^N,\,\dots,\,\Phi_{k-1}^N,\,z^2\,\Phi_{k}^N,\,\Phi_{k+1}^N,\,\dots,\,\Phi_{l-1}^N,\,z^2\,\Phi_{l}^N,\,\Phi_{l+1}^N,\dots \right|\\
 =8N(N-1)\left|\Phi_{0}^N,\, \Phi_1^N,\,\Phi_3^N,\dots \right|.
 \ee
Combining (\ref{High1})-(\ref{High4}) and an expression for the Virasoro operator $ \widehat{\mathsf{L}}_{2}=\widehat{Y}_p+\frac{\p^2}{\p t_2^2}$, we obtain
\be\label{Viruni}
 \widehat{\mathsf{L}}_{2} \, \tau_N=0.
\ee

To find a full algebra of the Virasoro constraints it is enough to consider nested commutators of the operators  $\widehat{\mathsf{L}}_{2}$ with $\widehat{\mathsf{L}}_{-1}$.
 The resulting operators 
 \be\label{Virall}
\widehat{\mathsf{L}}_{k}=\widehat{L}_{2k}-\frac{\p}{\p t_{2k+3}}+3N\frac{\p}{\p t_{2k}}+\sum_{j=1}^{k-1}\frac{\p^2}{\p t_{2j}\p t_{2k-2j}}+\left(\frac{1}{8}+\frac{3N^2}{2}\right)\delta_{k,0}+2Nt_2\delta_{k,-1},\,\,k\geq-1
 \ee
constitute an extension of the algebra (\ref{strdil}) to an infinite subalgebra of the Virasoro algebra 
\be
\left[\widehat{\mathsf{L}}_{k},\widehat{\mathsf{L}}_{m}\right]=2(k-m)\widehat{\mathsf{L}}_{k+m},
\ee
and annihilate the tau-function
\be\label{lprimec}
\widehat{\mathsf{L}}_{k} \tau_N =0, \,\,\,\,\,\,k\geq-1.
\ee

Let us now construct the operators $\widehat{\mathsf{M}}_{k}$ for $k>2$. 
One can find all higher $\widehat{\mathsf{M}}_{k}$ operators assuming that the commutation relation (\ref{smallcom}) holds for $k=-1,0,1$ and arbitrary $l$. Then, a commutation relation
\be
\left[\widehat{\mathsf{L}}_{-1},\widehat{\mathsf{M}}_{3}\right]=-10\widehat{\mathsf{M}}_{2}
\ee
allows us to find 
\be
\widehat{\mathsf{M}}_{3}=\widehat{M}_{6}-2\widehat{L}_{9}+\frac{\p}{\p t_{12}}+\left(12N^2+\frac{1}{4}\right)\frac{\p}{\p t_6}
+7N\left(\widehat{L}_{6}-\frac{\p}{\p t_9}\right)
+2N\frac{\p^2}{\p t_{2}\p t_{4}}-\frac{4}{3}\frac{\p^3}{\p t_{2}^3}.
\ee
Then, a commutation relation between the Virasoro and W-operators
\be
\left[\widehat{\mathsf{L}}_{1},\widehat{\mathsf{M}}_{l}\right]=2\,(2-l)\widehat{\mathsf{M}}_{l+1}
\ee
yields all operators $\widehat{\mathsf{M}}_{k}$ for $k\geq-2$
\be
\widehat{\mathsf{M}}_{k}=\widehat{M}_{2k}-2\widehat{L}_{2k+3}+\widehat{J}_{2k+6}+\left(3(k+1)N^2+\frac{1}{4}\right)\widehat{J}_{2k}\\
+(k+4)N\left(\widehat{L}_{2k}-\widehat{J}_{2k+3}\right)+2\left(N^2+\frac{1}{4}\right)N\delta_{k,0}+4\,N^2t_2\delta_{k,-1}+16\,N^2t_4\delta_{k,-2}\\
+(k-2)N\sum_{j=1}^{k-1}\frac{\p^2}{\p t_{2j}\p t_{2k-2j}}-\frac{4}{3}\sum_{i+j+l=k}\frac{\p^3}{\p t_{2i}\p t_{2j}\p t_{2l}}.
\ee
A straightforward calculation shows that the following commutation relations between the Virasoro and W-operators hold
\be
\left[\widehat{\mathsf{L}}_{k},\widehat{\mathsf{M}}_{l}\right]=2\,(2k-l)\widehat{\mathsf{M}}_{k+l}-4\left(k(k-1)-2\delta_{k,-1}\right)\,N\,\widehat{\mathsf{L}}_{k+l}+8\sum_{j=1}^{k-1}j \frac{\p}{\p t_{2k-2j}}\widehat{\mathsf{L}}_{l+j}
\ee
for $k\geq -1$ and $l\geq -2$, so that
\be\label{Wconstmaa}
\widehat{\mathsf{M}}_{k} \tau_N =0, \,\,\,\,\,\, k\geq-2.
\ee

Of course, one can choose another basis in the space of constraints. Let us consider the operators
\be
\widehat{\mathsf{M}}_{-2}' =\widehat{\mathsf{M}}_{-2},\\
\widehat{\mathsf{M}}_{k}' =\widehat{\mathsf{M}}_{k} -(k-2)\,N\,\widehat{\mathsf{L}}_{k},\,\,\,\,\,k>-2,
\ee
which also annihilate the tau-function
\be\label{mprimec}
\widehat{\mathsf{M}}_{k}' \tau_N =0, \,\,\,\,\,\,k\geq-2.
\ee
These operators satisfy the commutation relations
\be
\left[\widehat{\mathsf{L}}_{k},\widehat{\mathsf{M}}_{l}'\right]=2\,(2k-l)\widehat{\mathsf{M}}_{k+l}'+c_{kl}\,N\,\widehat{\mathsf{L}}_{k+l}+8\sum_{j=1}^{k-1}j \frac{\p}{\p t_{2k-2j}}\widehat{\mathsf{L}}_{l+j}\,\,\,\,\,\,\,\,\,\,\,\,\,\,\,\,k\geq -1,\,\,l\geq -2.
\ee
Here a combination of the Kronecker symbols $c_{kl}=8\left(\delta_{k,-1}(1-\delta_{l,-1})-(k+2)\delta_{l,-2}(1-\delta_{k,0})\right)$ guarantees that in the r.s.h.~there appear operators $\widehat{\mathsf{L}}_{k}$ only with $k\geq-1$.
In particular, we have
\be
\widehat{\mathsf{M}}_{-1}'=\widehat{M}_{-2}-2\widehat{L}_{1}+6N\left(\widehat{L}_{-2}-\frac{\p}{\p t_1}\right)+\frac{\p}{\p t_4}+\left(10N^2+\frac{1}{2}\right)t_2,\\
\widehat{\mathsf{M}}_{0}'=\widehat{M}_{0}-2\widehat{L}_{3}+6N\left(\widehat{L}_{0}-\frac{\p}{\p t_3}\right)+\frac{\p}{\p t_6}+\left(\frac{3}{4}+5N^2\right)N,\\
\widehat{\mathsf{M}}_{k}'=\widehat{M}_{2k}-2\widehat{L}_{2k+3}+6N\left(\widehat{L}_{2k}-\widehat{J}_{2k+3}\right)+\widehat{J}_{2k+6}+\left(9N^2+\frac{1}{4}\right)\widehat{J}_{2k}\\
-\frac{4}{3}\sum_{i+j+l=k}\frac{\p^3}{\p t_{2i}\p t_{2j}\p t_{2l}}\,\,\,\,\,\,\,\,\,\,\,\,\,\,\,\,k>0.
\ee

At the end of this section let us describe a simple Sugawara construction of the Virasoro constraints (\ref{Virall}). For this purpose we introduce the bosonic operators
\be\label{Sugast}
\widehat{\mathsf{J}}_k =
\begin{cases}
\displaystyle{\frac{\p}{\p t_k} \,\,\,\,\,\,\,\,\,\,\,\, \mathrm{for\,\,odd} \quad k>0},\\[8pt]
\displaystyle{\sqrt{3}\frac{\p}{\p t_k} \,\,\,\,\,\,\,\,\,\,\,\, \mathrm{for\,\,even} \quad k>0},\\[8pt]
\displaystyle{\sqrt{3}N}\,\,\,\,\,\,\,\,\,\,\,\,\,\,\,\,\,\,\, \mathrm{for} \quad k=0,\\[8pt]
\displaystyle{-kt_{-k} \,\,\,\,\,\mathrm{for \,\,odd} \quad k<0,}\\[8pt]
\displaystyle{-\frac{k}{\sqrt{3}}t_{-k} \,\,\,\,\,\mathrm{for\,\,even} \quad k<0,}
\end{cases}
\ee
and the corresponding bosonic current
\be\label{bosdef}
\widehat{\mathsf{J}}(z)=\sum_{k=-\infty}^\infty \left( \widehat{\mathsf{J}}_k-\delta_{k,-3}\right) z^{-k-1}
\ee
with the dilaton shift $\tilde{t}_k=t_k-\frac{1}{3}\delta_{k,3}$.\footnote{ Let us stress that the only difference between the standard bosonic current and (\ref{bosdef}) is the normalization of the even components. Thus, the current (\ref{bosdef}) can be reduced to the standard one by the change of even times $t_{2k}\mapsto \sqrt{3} t_{2k}$.} Then
\be
\frac{1}{2} \normordboson \widehat{\mathsf{J}}(z)^2 \normordboson=\sum_{k \in \z/2}
\frac{ \widehat{\mathsf{L}}_{k}}{z^{2k+2}}-\frac{1}{8z^2}.
\ee



\section{Cut-and-join operator for general $N$}\label{S5}

Following the idea of \cite{MorSh} in this section we construct the cut-and-join operator representation for the tau-function $\tau_N$. 
Let us introduce the gradation $\deg t_k =\frac{k}{3}$ such, that
\be
\deg{\widehat J}_k=\deg \widehat{L}_k=\deg{\widehat{M}_k}=-\frac{k}{3},
\ee
and the degree operator
\be
\widehat{D}=\frac{1}{3}\sum_{k=1}^\infty k t_k \frac{\p}{\p t_k}.
\ee
 Then, the operators
 \be
\widehat{\mathsf{L}}_{k}^*=\widehat{\mathsf{L}}_{k}+\frac{\p}{\p t_{2k+3}},\,\,\,\,k\geq -1,
\ee
have the degree $-2k/3$, and the operators
\be
\widehat{\mathsf{M}}_{k}^*=\widehat{\mathsf{M}}_{k}'-\frac{\p}{\p t_{2k+6}},\,\,\,\,k\geq -2,\\
\ee
consist of the terms with degree $-2k/3$ and $-(2k/3+1)$. From the Virasoro  and W-constraints  (\ref{lprimec}) and (\ref{mprimec}) it immediately follows that
\be\label{Wmasteq}
\widehat D\,  \tau_N =\frac{1}{3}\sum_{k=0}^\infty\left( (2k+1) t_{2k+1}\widehat{\mathsf{L}}_{k-1}^*-(2k+2) t_{2k+2}\widehat{\mathsf{M}}^*_{k-2} \right) \tau_N.
\ee 
An operator in the r.h.s. is a sum of the operators $\widehat{\mathsf{W}}_{1}$ and $\widehat{\mathsf{W}}_{2}$:
\be\label{cajgen}
\widehat{\mathsf{W}}_{1}=\frac{1}{3}\left(\sum_{k=0}^\infty k t_k \frac{3+(-1)^k}{2}\left(\widehat{L}_{k-3}+3N\frac{\p}{\p t_{k-3}}\right)\right.\\
\left.+\sum_{k=3}^\infty(2k+1)(k+1)t_{2k+1}\sum_{j=1}^{k-2}\frac{\p^2}{\p t_{2j}\p t_{2(k-j-1)}}+3\left(\frac{3N^2}{2}+\frac{1}{8}\right)t_3+6Nt_1t_2\right),\\

\widehat{\mathsf{W}}_{2}=-\frac{2}{3}\left(\sum_{k=0}^\infty(k+1)t_{2k+2}\left(\widehat{M}_{2k-4}+6N\widehat{L}_{2k-4}\right)-4Nt_2\widehat{L}_{-4}\right.\\
\left.+\sum_{k=3}^\infty(k+1)t_{2k+2}\left(\left(9N^2+\frac{1}{2}\right)\frac{\p}{\p t_{2k-4}}-\frac{4}{3}\sum_{i+j+l=k-1}\frac{\p^3}{\p t_{2i}\p t_{2j}\p t_{2l}}
\right)\right.\\
\left.+2(12N^2+1)t_2t_4+\left(\frac{9}{4}+15N^2\right)t_6\right)
\ee
such that
\be
\deg{\widehat{\mathsf{W}}_{1}}=1, \,\,\,\,\,\deg{\widehat{\mathsf{W}}_{2}}=2.
\ee
From (\ref{Wmasteq}) it is clear that $\tau_N$ is a sum of components with integer degree:
\be
\tau_N=1+\sum_{k=1}^\infty \tau_N^{(k)},
\ee
where $\deg \tau_N^{(k)}=k$. Let us introduce a variable $q$ which counts the degree:
\be
\tau_N(q)=1+\sum_{k=1}^\infty \tau_N^{(k)}q^k.
\ee
Then the operator $\widehat{D}$ acts as a derivative $q \frac{\p}{\p q}$, so that $\tau_N(q)$ satisfies the cut-and-join type equation
\be
q\frac{\p}{\p q} \tau_N(q)=\left(q\widehat{\mathsf{W}}_{1}+q^2\widehat{\mathsf{W}}_{2}\right) \tau_N(q).
\ee
For the commuting operators $\widehat{\mathsf{W}}_{1}$ and $\widehat{\mathsf{W}}_{2}$ the solution would be 
\be
\exp \left(q\widehat{\mathsf{W}}_{1}+\frac{q^2}{2}\widehat{\mathsf{W}}_{2}\right) \cdot1,
\ee
but it is easy to check that
\be
\left[\widehat{\mathsf{W}}_{1},\widehat{\mathsf{W}}_{2}\right]\neq 0.
\ee
Thus, the solution can be represented in terms of an ordered exponential, and the operators (\ref{cajgen}) define a recursion
\be
\tau_N^{(k)}=\frac{1}{k}\left(\widehat{\mathsf{W}}_{1}\,\tau_N^{(k-1)}+\widehat{\mathsf{W}}_{2}\,\tau_N^{(k-2)}\right),
\ee
with the initial conditions $\tau_N^{(0)}=1$,  $\tau_N^{(-1)}=0$. In particular
\be
\tau_N^{(1)}=\widehat{\mathsf{W}}_{1} \cdot 1,\\
\tau_N^{(2)}=\frac{1}{2}\left(\widehat{\mathsf{W}}_{1}^2+\widehat{\mathsf{W}}_{2} \right)\cdot 1,\\
\tau_N^{(3)}=\frac{1}{3!}\left(\widehat{\mathsf{W}}_{1}^3+\widehat{\mathsf{W}}_{1}\widehat{\mathsf{W}}_{2}+2\widehat{\mathsf{W}}_{2}\widehat{\mathsf{W}}_{1} \right)\cdot 1.
\ee
Explicit expression for these three terms can be found in Appendix A.

From our construction it is clear that the operators $\widehat{\mathsf{W}}_{1,2}$ are not unique. In particular, if one substitutes the operator $\widehat{\mathsf{M}}^*_{k}$ in (\ref{Wmasteq}) with an operator  $\widehat{\mathsf{M}}^*_{k}+\beta_k \mathsf{L}_k$
for arbitrary constant $\beta_k$'s the equation remains valid. This gives the following change in the operators:
\be
\Delta\widehat{\mathsf{W}}_{1}=\frac{2}{3}\sum_{k=2}^\infty k\,\beta_{k-3}\,t_{2k}\frac{\p}{\p t_{2k-3}},\\
\Delta\widehat{\mathsf{W}}_{2}=-\frac{2}{3}\sum_{k=2}^\infty k\,\beta_{k-3}\,t_{2k}\widehat{\mathsf{L}}_{k-3}^*.
\ee
However, it is easy to see that this freedom is not enough to make the operators $\widehat{\mathsf{W}}_{1}$ and $\widehat{\mathsf{W}}_{2}$ commuting with each other. However, one can consider more general transformations, generated by the operators $\sum_{j=-1}^\infty\beta_{k,j}\widehat{J}_{k-j}\widehat{\mathsf{L}}_{j}$ with some constant matrix $\beta_{k,j}$. It is not clear if this is enough to make the operators commutative.

\section{Open intersection numbers: Virasoro and W-constraints}\label{S6}

In this section we consider the case $N=1$ which, as we proved in Section \ref{S2}, describes open intersection numbers of \cite{PST, Buryak, Buryak2}.

As we have established in \cite{Aopen}, an operator
\be
a_{1}=z \,a_{KW}\, z^{-1}=\frac{1}{z}\frac{\p}{\p z} -\frac{3}{2}\frac{1}{z^2}+z 
\ee
is the KS operator for the tau-function $\tau_1$. Moreover, in this case we have a family of the KS operators
\be
l_k^o=-z^{2k+2}a_1, \,\,\,\,\, k\geq -1,
\ee
 which constitute a subalgebra of the Virasoro algebra and guarantee \cite{Aopen} that the tau-function satisfies the Virasoro constraints
 \be\label{Virconst}
 \widehat{L}_k^o \,\tau_1 =0 ,\,\,\,\,k>-1.
 \ee
 Here
 \be
 \widehat{L}_{k}^o=\widehat{L}_{2k}+ (k+2)\widehat{J}_{2k}-\widehat{J}_{2k+3}+\delta_{k,0}\left(\frac{1}{8}+\frac{3}{2}\right).
 \ee
Thus, for $N=1$ we have two sets of the Virasoro constraints, namely (\ref{Viruni}) and  (\ref{Virconst}), which do not coincide for $k>1$. The difference is
\be
\widehat{O}_k= \widehat{L}_{k}^o-\left.\widehat{\mathsf{L}}_{k}\right|_{N=1}=(k-1)\frac{\p}{\p t_{2k}}-\sum_{j=1}^{k-1}\frac{\p^2}{\p t_{2j}\p t_{2(k-j)}}.
\ee
From the constraints 
\be
\widehat{O}_k \tau_1=0,\,\,\,\,k>1,
\ee
we obtain the relations, which describe the dependence of the tau-function on even times $t_{2k}$ for $k>1$,\footnote{These constraints immediately follow from the residue formula (\ref{Resid}). Let us note that the same relations are true for the tau-function $\tau$ given by (\ref{genres}), for any $\Phi_1$ if $\widetilde{\tau}$ is an arbitrary tau-function of the KdV hierarchy.}
\be
\frac{\p}{\p t_{2k}}\tau_1=\frac{\p^k}{\p t_2^k}\tau_1.
\ee
This property of the generating function of open intersection numbers has been established in \cite{Buryak2}. This equation allows us to describe a dynamics with respect to the times $t_{2k}$ for $k>1$:
\be
\tau_1({\bf t})=\exp{\left(\sum_{k=2}^\infty t_{2k} \frac{\p^k}{\p t_{2}^k}\right)}\,\tau_1(t_1,t_2,t_3,0,t_5,0,t_7,0,\dots).
\ee

Thus, there is a one-parametric family of the constraints
\be
 \widehat{L}_{k}^o(\alpha)=\widehat{L}_{k}^o+\alpha\widehat{O}_k,
\ee
where we assume that $\widehat{O}_k=0$ for $k=-1,0,1$. These operators satisfy the Virasoro commutation relations
\be
\left[\widehat{L}_k^o(\alpha),\widehat{L}_m^o(\alpha)\right]=(k-m)\widehat{L}_{k+m}^o(\alpha),
\ee 
and annihilate the tau-function $\tau_1$. The Virasoro constraints obtained in \cite{Buryak2} correspond to $\alpha=1/2$.

In addition to the Virasoro constraints we have infinitely many higher W-constraints. Let us consider the KS operators
 \be
w_k^o=z^{2k+4}a_1^2,\,\,\,\, k\geq-2.
\ee
They satisfy the following commutation relations 
\be
\left[w_k^o,l_m^o\right]=2(k-2m)w_{k+m}^o+4m(m+1)l_{m+k}^o,
\ee
while commutators $[w_k,w_l]$ contain terms with an operator $a_1^3$. Using the correspondence (\ref{Mastid}) we construct the following operators from $W_{1+\infty}$:
\be
\widehat{M}_{-2}^o=\widehat{M}_{-4}+2\widehat{L}_{-4}-2\widehat{L}_{-1}+5t_4-2t_1+\frac{\p}{\p t_2},\\
\widehat{M}_{-1}^o=\widehat{M}_{-2}+4\widehat{L}_{-2}-2\widehat{L}_{1}+\frac{13}{2}t_2-4\frac{\p}{\p t_1}+\frac{\p}{\p t_4},\\
\widehat{M}_{0}^o=\widehat{M}_{0}+6\widehat{L}_{0}-2\widehat{L}_{3}-6\frac{\p}{\p t_3}+\frac{\p}{\p t_6}+\frac{23}{4},\\
\widehat{M}_{k}^o=\widehat{M}_{2k}+2(k+3)\widehat{L}_{2k}-2\widehat{L}_{2k+3}+\left(\frac{95}{12}+6k+\frac{4}{3}k^2\right)\frac{\p}{\p t_{2k}}\\
-2(k+3)\frac{\p}{\p t_{2k+3}}+\frac{\p}{\p t_{2k+6}},\,\,\,\,\,\,\,\, k\geq 1,
\ee
or
\be
\widehat{M}_k^o=\widehat{M}_{2k}+2(k+3)\widehat{L}_{2k}-2\widehat{L}_{2k+3}-2(k+3)\widehat{J}_{2k+3}\\
+\left(\frac{95}{12}+6k+\frac{4}{3}k^2\right)\widehat{J}_{2k}+\widehat{J}_{2k+6}+\frac{23\,\delta_{k,0}}{3}.
\ee
The constant term in $\widehat{\mathsf{M}}_{0}^o$ is chosen in such a way that the following commutation relations hold
\be
\left[\widehat{M}_k^o,\widehat{L}_m^o\right]=2(k-2m)\widehat{M}_{k+m}^o+4m(m+1)\widehat{L}_{k+m}^o,\,\,\,\,\,\,\, k\geq-2,m\geq-1.
\ee
These commutation relations guarantee that
\be\label{Wconst}
\widehat{M}_{k}^o\, \tau_1=0\,\,\,\, k\geq -2.
\ee

\section{Cut-and-join operator for open intersection numbers}\label{S7}

In Section \ref{S5} we have already obtained the cut-ant-join operator description of the Kontsevich--Penner model, which is valid for arbitrary $N$. In particular, it is valid for the case $N=1$, corresponding to open intersection numbers. However, as we have seen in the previous section, for this case the Virasoro and W-constraints have additional parameters, thus, there is a vast family of the cut-and-join type operators. Here we construct a representative of this family. We will directly follow the construction of Section  \ref{S5}. Let us introduce the shifted operators
\be
\widehat{L}_{k}^*=\widehat{L}_{k}^o+\frac{\p}{\p t_{2k+3}},\,\,\,\,k\geq -1,
\ee
and
\be
\widehat{M}_{-2}^*=-\widehat{M}_{-2}^o+\frac{\p}{\p t_{2}},\\
\widehat{M}_{k}^*=-\widehat{M}_{k}^o+2(k+3)\widehat{L}_{k}^o+\frac{\p}{\p t_{2k+6}}\\
=-\widehat{M}_{2k}+2\widehat{L}_{2k+3}
+\left(\frac{49}{12}+4k+\frac{2}{3}k^2\right)\widehat{J}_{2k}+4\delta_{k,0},\,\,\,\,k\geq-1
\ee
so that
\be
\widehat{L}_{k}^*\,\tau_1=\frac{\p}{\p t_{2k+3}}\,\tau_1, \,\,\,\,k\geq-1,\\
\widehat{M}_{k}^*\,\tau_1=\frac{\p}{\p t_{2k+6}}\,\tau_1,\,\,\,\, k\geq-2.
\ee
Then
\be\label{grd}
\widehat D\,  \tau_1 =\frac{1}{3}\sum_{k=0}^\infty\left( (2k+1) t_{2k+1}\widehat{L}_{k-1}^*+(2k+2) t_{2k+2}\widehat{M}^*_{k-2} \right) \tau_1.
\ee
Again, the operator in the r.h.s. of (\ref{grd}) is a sum of two operators
\be
\widehat{{W}}_{1}^o=\frac{1}{3}\left(\sum_{k=0}^\infty k t_k \frac{3+(-1)^k}{2}\widehat{L}_{k-3}+\sum_{k=2}^\infty(2k+1)(k+1)t_{2k+1}\frac{\p}{\p t_{2k-2}}+\frac{39}{8}t_3+6t_1t_2\right),\\
\widehat{{W}}_{2}^o=\frac{2}{3}\left(-\sum_{k=0}^\infty(k+1)t_{2k+2}\widehat{M}_{2k-4}-2t_2\widehat{L}_{-4}\right.\\
\left.+\sum_{k=3}^\infty(k+1)\left(\frac{2}{3}k^2+\frac{4}{3}k-\frac{5}{4}\right)t_{2k+2}\frac{\p}{\p t_{2k-4}}-2t_2t_4+12t_6\right),
\ee
such that
\be
\deg{\widehat{{W}}_{1}^o}=1, \,\,\,\,\,\deg{\widehat{W}^o_{2}}=2.
\ee
For the expansion of the tau-function
\be
\tau_1=1+\sum_{k=1}^\infty \tau_1^{(k)},
\ee
where $\deg \tau_1^{(k)}=k$, we have a recursion
\be
\tau_1^{(k)}=\frac{1}{k}\left(\widehat{{W}}_{1}^o\tau_1^{(k-1)}+\widehat{{W}}_{2}^o\tau_1^{(k-2)}\right)
\ee
such that
\be
\tau_1^{(1)}=\widehat{{W}}_{1}^o \cdot 1,\\
\tau_1^{(2)}=\frac{1}{2}\left(\left(\widehat{{W}}_{1}^o\right)^2+\widehat{{W}}_{2}^o \right)\cdot 1,\\
\tau_1^{(3)}=\frac{1}{3!}\left(\left(\widehat{{W}}_{1}^o\right)^3+\widehat{{W}}_{1}^o\,\widehat{{W}}_{2}^o+2\widehat{{W}}_{2}^o\,\widehat{{W}}_{1}^o \right)\cdot 1.
\ee
These three terms give the expansion of $\tau_o$, presented in \cite{Aopen}.






\section{Other interesting values of $N$}\label{S8}

In general, for integer $N>0$, operators $z^{2k}(a_N)^{N+m}$ with non-negative $m$ and $k$ are the KS operators for the tau-function $\tau_N$. Indeed, an operator $(a_N)^N$ maps the basis vectors (\ref{bascon}) into the set, proportional to the basis vectors of the KW model
\be
\left(a_N\right)^N \left\{\Phi^N\right\}=z^{N}\left\{\Phi^{KW}\right\} 
\ee
and, on this space, operators $z^2$ and $a_N$ are the KS operators.
Thus, for integer $N>1$ we have additional constraints, which follow from the existence of two different families of the $W^{(N+1)}$ constraints. These constraints allow us to restore the dependence of the tau-function on the even times $t_{2k}$ for $k>N$ from the dependence of the times $t_2,\dots,t_{2N}$. 

For example, for $N=2$ the operators $z^{4+2k}a_N^2$ for $k\geq-2$ are the KS operators. Thus, for $N=2$ in addition to the constraints (\ref{Wconstmaa}) we have the operators
\be
\widehat{M}_{k}''=\widehat{M}_{2k}+2(k+4)\left(\widehat{L}_{2k}-2\widehat{J}_{2k+3}\right)
+\widehat{J}_{2k+6}\\
+\left(\frac{4}{3}k^2+8k+\frac{179}{12}\right)\widehat{J}_{2k}
-\widehat{L}_{2k+3}+17\,\delta_{k,0},\,\,\,\,k\geq-2,
\ee
which annihilate the tau-function
\be
\widehat{M}_{k}''\, \tau_2=0.
\ee
The simplest equation, which follows from the existence of two different families of the constraints, is
\be
\frac{\p}{\p t_6}\tau_2({\bf t})=\widehat{P}_3\, \tau_2({\bf t}),
\ee
where
\be
\widehat{P}_3=\frac{3}{2}\frac{\p^2}{\p t_2\p t_4}-\frac{1}{2}\frac{\p^3}{\p t_2^3}.
\ee
This equation describes a dynamics with respect to the time $t_6$:
\be
\tau_2(t_1,t_2,t_3,t_4,t_5,t_6,t_7,\dots)=\exp{\left(t_6 \widehat{P}_1\right)}\,\tau_2(t_1,t_2,t_3,t_4,t_5,0,t_7,\dots).
\ee
In general, for $k>2$ we have
\be
\frac{\p}{\p t_{2k}}\tau_2({\bf t})=\widehat{P}_{k} \tau_2({\bf t}),
\ee
where 
\be
\widehat{P}_{k}=\frac{3}{2(k-1)}\sum_{j=1}^{k-1}\frac{\p^2}{\p t_{2j}\p t_{2k-2j}}-\frac{1}{(k-1)(k-2)}\sum_{i+j+l=k}\frac{\p^3}{\p t_{2i}\p t_{2j}\p t_{2l}}
\ee
describes the dependence on the time $t_{2k}$ (and, for arbitrary $k>2$, the operator $\widehat{P}_{k}$ can be reduced to the operator, acting only on the times $t_2$ and $t_4$). This dependence follows from the integral representation
\be
\tau_2({\bf{t}})=\oint\oint \left(\frac{1}{z_2}-\frac{1}{z_1}\right)e^{\sum_{k=1}^\infty  t_k ( z_1^k+z_2^k)}
{\tau}_{KW}\left({\bf t}-\left[z_1^{-1}\right]-\left[z_2^{-1}\right]\right) \Phi^2_1(z_1)\Phi^2_2(z_2)  \frac{d z_1}{2 \pi i z_1} \frac{d z_2}{2 \pi i z_2}
\ee
which is a particular case of (\ref{GENI}).

 \section*{Acknowledgments}
The author is grateful to A. Buryak,  A. Mironov,  A. Morozov and R. Tessler  for useful discussions. This work was supported 
in part by RFBR grant 14-01-00547 and NSh-1500.2014.2. The author would also like to thank
the hospitality of the ETH Zurich, where this work was completed.

\newpage
\appendix
\section*{Appendix A: Explicit formulas for $\tau_N$ and $\log(\tau_N)$}
\label{schur}
\def\theequation{A\arabic{equation}}
\def\theHequation{\theequation}
\setcounter{equation}{0}

\be
\tau_N=1+ \left( \frac{1}{8}+\frac{3}{2}\,{N}^{2} \right) t_{{3}}+\frac{1}{6}\,{t_{{1}}}^{3}+2N\,t_{{1
}}t_{{2}}
+\left( 3\,{N}^{3}+{\frac {25}{4}}\,N \right) t_{{1}}t_{{3}}t_{{2}}+
4N \left( 1+{N}^{2} \right) t_{{6}}\\
 +\frac{4}{3}\,N{t_{{2}}}^{3}+ \left( {
\frac {9}{8}}\,{N}^{4}+{\frac {39}{16}}\,{N}^{2}+{\frac {25}{128}}
 \right) {t_{{3}}}^{2}
 +{\frac {1}{72}}\,{t_{{1}}}^{6}+\frac{1}{3}\,N{t_{{1}}}^
{4}t_{{2}}+ \left(\frac{1}{4}\,{N}^{2}+{\frac {25}{48}} \right) t_{{3}}{t_{{1
}}}^{3}
+8\,{N}^{2}t_{{2}}t_{{4}}\\
+4\,N{t_{{1}}}^{2}t_{{4}}+2\,{N}^{2}{t
_{{1}}}^{2}{t_{{2}}}^{2}+ \left(\frac{15}{2}\,{N}^{2}+\frac{5}{8} \right) t_{{1}}t_{{
5}}+{\frac {28}{3}}\,{N}^{2}{t_{{1}}}^{3}t_{{2}}t_{{4}}+32\,t_{{1}}{N}^{2}
{t_{{4}}}^{2}+ 20\,N\left( 1+2\,{N}^{2} \right) t_{{5}}t_{{4}}\\
+ 32\,N\left( 1+{N}^{2} \right) t_{{1}}t_{{8}}
 + \left( {\frac {35}{
16}}+{\frac {105}{4}}\,{N}^{2} \right) {t_{{1}}}^{2}t_{{7}}+35N \left( {
\frac {3}{4}}+{N}^{2} \right) t_{{2}}t_{{7}}+ \left( \frac{2}{9}\,N+\frac{4}{3}
\,{N}^{3} \right) {t_{{1}}}^{3}{t_{{2}}}^{3}\\
+ \left( {\frac {49}{6}}
\,N+2\,{N}^{3} \right) {t_{{2}}}^{3}t_{{3}}
+ \left( \frac{5}{4}\,{N}^{2}+{
\frac {35}{48}} \right) {t_{{1}}}^{4}t_{{5}}+ \left( \frac{3}{16}\,{N}^{4}+{
\frac {1225}{768}}+{\frac {37}{32}}\,{N}^{2} \right) {t_{{1}}}^{3}{t_{
{3}}}^{2}\\
+ \left( {\frac {26}{3}}\,N+\frac{2}{3}\,{N}^{3} \right) {t_{{1}}}^{3
}t_{{6}}
+ \left( 6\,{N}^{5}+{\frac {61}{2}}\,{N}^{3}+{\frac {49}{2}}\,
N \right) t_{{6}}t_{{3}}+ \frac{1}{48}\left({N}^{2}+{\frac {49}{12}}
 \right) {t_{{1}}}^{6}t_{{3}}\\+ \left( {\frac {105}{8}}\,{N}^{4}+{
\frac {735}{16}}\,{N}^{2}+{\frac {105}{128}} \right) t_{{9}}
+ \left( {
\frac {1225}{3072}}+{\frac {9}{16}}\,{N}^{6}+{\frac {1299}{256}}\,{N}^
{2}+{\frac {225}{64}}\,{N}^{4} \right) {t_{{3}}}^{3}+{\frac {1}{1296}}
\,{t_{{1}}}^{9}\\
+\frac{1}{36}\,N{t_{{1}}}^{7}t_{{2}}+\frac{2}{3}\,N{t_{{1}}}^{5}t_{{4}
}+\frac{8}{3}\,{N}^{2}t_{{1}}{t_{{2}}}^{4}+ \left( {\frac {245}{64}}+{\frac {
375}{8}}\,{N}^{2}+{\frac {45}{4}}\,{N}^{4} \right) t_{{1}}t_{{3}}t_{{5
}}\\
+ \left( {\frac {49}{2}}\,N+6\,{N}^{3} \right) {t_{{1}}}^{2}t_{{3}}t
_{{4}}+ \left( 49\,{N}^{2}+12\,{N}^{4} \right) t_{{2}}t_{{3}}t_{{4}}+
 \left( {\frac {111}{8}}\,{N}^{3}+\frac{9}{4}\,{N}^{5}+{\frac {1225}{64}}\,N
 \right) t_{{2}}t_{{1}}{t_{{3}}}^{2}\\
 + \left( 56\,{N}^{2}+8\,{N}^{4}
 \right) t_{{2}}t_{{1}}t_{{6}}+ \left( \frac{1}{2}\,{N}^{3}+{\frac {49}{24}}\,
N \right) t_{{2}}{t_{{1}}}^{4}t_{{3}}+ 16N\left( 1+{N}^{2}
 \right) t_{{1}}{t_{{2}}}^{2}t_{{4}}\\
 + \left( 3\,{N}^{4}+{\frac {49}{4}
}\,{N}^{2} \right) {t_{{1}}}^{2}{t_{{2}}}^{2}t_{{3}}+ \left( {\frac {
65}{4}}\,N+15\,{N}^{3} \right) t_{{2}}{t_{{1}}}^{2}t_{{5}}+30\,{N}^{2}{t_{{2}}}^{2}t_{{5}}+\frac{1}{3}\,{N}^{2}{t_{{1}}}^
{5}{t_{{2}}}^{2}+\dots
\ee

\be
\log(\tau_N)= \left(\frac{1}{8}+\frac{3}{2}\,{N}^{2} \right) t_{{3}}+\frac{1}{6}\,{t_{{1}}}^{3}+2N\,t_{{1}}t_{{2}}
+6\,Nt_{{1}}t_{{2}}t_{{3}}
+4N \left(1+{N}^{2} \right) t_{{6}}\\+\frac{4}{3}
\,N{t_{{2}}}^{3}+ \left( \frac{9}{4}\,{N}^{2}+\frac{3}{16} \right) {t_{{3}}}^{2}+\frac{1}{2}\,
t_{{3}}{t_{{1}}}^{3}+8\,{N}^{2}t_{{2}}t_{{4}}
+4\,{t_{{1}}}^{2}Nt_{{4}}
+ \left( \frac{15}{2}\,{N}^{2}+\frac{5}{8} \right) t_{{1}}t_{{5}}\\
+8\,N{t_{{2}}}^{3}t_{{3}}+ 15\left( 3\,{N}^{2}+{\frac {1}{4}} \right) t
_{{1}}t_{{3}}t_{{5}}+24\,N{t_{{1}}}^{2}t_{{3}}t_{{4}}+30\,{N}^{2}{t_{{
2}}}^{2}t_{{5}}+ \frac{105}{8}\left( {\frac {1}{16}}+{\frac {7}{2}}\,{N}^{2}+
{N}^{4} \right) t_{{9}}\\
+ 35\left( {N}^{3}+{\frac {
3}{4}}\,N \right) t_{{7}}t_{{2}}+ 35\left( {\frac {1}{16}}+{\frac {
3}{4}}\,{N}^{2} \right) t_{{7}}{t_{{1}}}^{2}+ 32\,N\left( {N}^{
2}+1\right) t_{{8}}t_{{1}}+32{N}^{2}\,t_{{1}}{t_{{4}}}^{2}\\
+48\,{N}^{2}t
_{{2}}t_{{3}}t_{{4}}+18\,Nt_{{1}}t_{{2}}{t_{{3}}}^{2}+ 20\,N\left( 1+2
\,{N}^{2} \right) t_{{5}}t_{{4}}+ 24N\left({N}^{2}+1 \right) t_{
{6}}t_{{3}}+8\,N{t_{{1}}}^{3}t_{{6}}\\
+\frac{3}{2}\,{t_{{1}}}^{3}{t_{{3}}}^{2}+
48\,{N}^{2}t_{{1}}t_{{2}}t_{{6}}+16\,Nt_{{1}}{t_{{2}}}^{2}t_{{4}}+\frac{5}{8}
\,{t_{{1}}}^{4}t_{{5}}+ \left(\frac{9}{2}\,{N}^{2}+\frac{3}{8} \right) {t_{{3}}}^{3}+
15\,N{t_{{1}}}^{2}t_{{2}}t_{{5}}+\dots
\ee

\end{document}